\documentclass[
tightenlines,
aps,
twocolumn,
nofootinbib,
floatfix]{revtex4}

\usepackage{graphicx}
\usepackage{float}
\usepackage{amsmath}

\usepackage{color}

\newcommand{\be}{\begin{equation}}
\newcommand{\ee}{\end{equation}}
\newcommand{\ba}{\begin{eqnarray}}
\newcommand{\ea}{\end{eqnarray}}

\def\vec#1{{\mbox{\boldmath$#1$}}}

\newcommand{\ep}{\epsilon}

\begin{document}
\begin{titlepage}

\begin{flushright}
\vbox{
\begin{tabular}{l}
ANL-HEP-PR-11-67\\
NUHEP-TH/11-24
\end{tabular}
}
\end{flushright}
\vspace{0.1cm}

\title{
A subtraction scheme for NNLO computations
}

\author{Radja Boughezal}
\email[]{rboughezal@hep.anl.gov}
\affiliation{High Energy Physics Division, Argonne National Laboratory, Argonne, IL 60439, USA} 

\author{Kirill Melnikov }
\email[]{melnikov@pha.jhu.edu}
\affiliation{Department of Physics and Astronomy, Johns Hopkins University, Baltimore, MD 21218, USA}

\author{Frank Petriello}
\email[]{f-petriello@northwestern.edu}
\affiliation{Department of Physics \& Astronomy, Northwestern University, Evanston, IL 60208, USA \\
High Energy Physics Division, Argonne National Laboratory, Argonne, IL 60439, USA}

\begin{abstract}
\vspace{2mm}
We use the known soft and collinear limits of 
tree- and one-loop 
scattering amplitudes -- computed over a decade ago -- to explicitly 
construct a subtraction scheme for next-to-next-to-leading order (NNLO) computations.  Our approach combines partitioning of the final-state 
phase space together with the technique of sector decomposition, following recent suggestions in Ref.~\cite{Czakon:2010td}.  We apply this 
scheme to a toy example: the NNLO QED corrections to the decay 
of the $Z$ boson to a pair of
massless leptons.  We argue that the main features of this subtraction 
scheme remain valid for computations of  processes of arbitrary complexity 
with NNLO accuracy. 
\end{abstract}

\maketitle

\thispagestyle{empty}
\end{titlepage}

\section{Introduction} 
Asymptotic freedom and factorization of short- and long-distance effects 
in the hard scattering of hadrons together allow us to employ 
perturbative computations in QCD to extract information about the
physics of parton-parton interactions.  Perturbative QCD computations are organized by ``loop order'' in an
expansion in the strong coupling constant.  At both 
leading- and next-to-leading order in $\alpha_s$, the conceptual framework 
for perturbative 
QCD computations is well-established. While this does not necessarily 
make such calculations easy, 
 or even feasible in some cases, the existence of a general 
framework is needed to claim
a full understanding  of the structure of perturbative QCD at these orders.

The crucial element of such an understanding is the concept of infrared 
and collinear safety. It leads to computational algorithms~\cite{Catani:1996vz} that permit the calculation 
of arbitrary hadron-collider observables to next-to-leading order (NLO) accuracy.  These methods rely upon 
approximating matrix elements containing an additional massless particle by appropriate limits in singular regions of phase space.  These limits are simple 
enough to be integrated over the unresolved phase space of the radiated particle, without any reference 
to a particular observable.  Unfortunately,  the situation is much more confusing at next-to-next-to-leading order (NNLO) and beyond.  To illustrate 
this point, we note that  no NNLO computation of a 
cross section for a $ 2 \to 2 $ scattering process with strongly interacting  
particles exists, in spite of the fact that 
a large number of two-loop virtual amplitudes   
\cite{bern1,babis1,babis2,glover,garland}
and {\it all} singular limits of tree- and one-loop amplitudes 
\cite{Catani:2000vq, Catani:1999ss,Kosower:1999rx}
have been known for over ten years. The reason is that 
a working algorithm that combines these ingredients to obtain physical cross sections has not been  formulated. 
Consequently, a  significant number of existing fully differential 
NNLO computations  \cite{an1,m1,cat1,cat2,an2,an3,an4,g1,g2,Anastasiou:2005pn,Biswas:2009rb,Melnikov:2008qs,arXiv:1107.1164,arXiv:1110.2368,arXiv:1110.2375}
 were performed using unorthodox approaches, that 
are only remotely related to 
mainstream NLO subtractions ideas considered generalizable to higher orders~\cite{Catani:1996vz}.

An alternative NLO subtraction scheme
due to  Frixione, Kunszt and Signer (FKS)~\cite{Frixione:1995ms},  was not the scheme of choice 
when NLO computations were initially being undertaken, 
but has received renewed interest recently~\cite{arXiv:0908.4272}.
The main idea  
of the FKS subtraction  is to partition  the phase-space such that in every sector only one definite external particle $i$ 
can become soft,  or two definite external particles $i$ and $j$ 
can become collinear to each other. If such a partitioning exists,  
it is  clear that all singularities in a given  sector  
are easily extracted if the unresolved phase space is parameterized in terms 
of the energy of particle $i$  and the relative angle between directions 
of particles $i$ and $j$. 

In the case of NNLO computations the ``elementary building block'' 
is the double-unresolved phase space, where two given particles can become 
soft or collinear to a third particle. Extraction of singularities 
in the triple collinear limit is non-trivial, but can be accomplished 
by applying  the  technique of sector decomposition 
\cite{binothheinrich1,binothheinrich2,an1}  to the three-parton 
unresolved phase-space.  However,  early 
applications of sector decomposition at NNLO~\cite{an1,an2,an3} did not perform an initial partition of the 
phase space to separate collinear singularities, and instead attempted to find a suitable phase-space 
parameterization for an {\it entire} process at NNLO.  The complexity 
of this endeavor slowed down the progress of NNLO computations for hadron-collider observables using sector decomposition after initial success with such computations 
for the hadro-production
of the Higgs and  electroweak bosons. 

It was recently pointed out by Czakon~\cite{Czakon:2010td,Czakon:2011ve} that by combining the idea of 
phase-space partitioning from FKS
with the idea that  sector 
decomposition can be applied to real emission integrals 
\cite{an1,binothheinrich2}, a powerful framework for NNLO computations is obtained.  The purpose of this 
paper is to elaborate on this observation and show explicitly how 
this framework can be used to obtain physics results. In addition,  we show how to handle collinear singularities in the final state,
and discuss in detail the computation of real-virtual corrections and the scheme dependence
of the results.  For the sake 
of simplicity, we study two-loop QED corrections 
to the decay rate of the $Z$-boson to an electron-positron pair.
Dealing with QED corrections offers  significant simplifications, 
yet is far from trivial, and is a good place in which to develop and test ideas.
 
In the following sections we explain in detail our approach. In Section II, we discuss the extraction of singularities from double-real 
radiation corrections. In Section III, we explain how one-loop 
corrections to the one-photon  real-emission process 
are treated. In Section IV we describe the dependence of the results 
on the choice of the regularization scheme. In Section V we present our 
conclusions. 

\section{Double real radiation} 
\label{sec:doublereal}

We consider the decay of a $Z$-boson to an $e^+e^-\gamma \gamma$ final state, 
$ Z(p_Z) \to e^+(p_+) + e^-(p_-) + \gamma(p_1) + \gamma(p_2) $. 
We must design a strategy to integrate the squared matrix element 
for this process over the phase space of the final-state particles. 
This is non-trivial because such an integration eventually leads to  phase-space
regions that contain collinear\footnote{We treat all final state particles 
as massless.}  and soft singularities.  The main problem is not that such singularities exist, but that in different regions of phase space, different 
subsets of particles lead to singularities in the matrix elements. As a first 
step, we partition the phase-space such that we know which final-state particles can develop singularities in each partition. 

We begin by describing the collinear partitioning.  We introduce four functions 
\be
\Delta_i^{\pm} = 1 - {\vec n}^\pm \cdot \vec n_i, 
\ee
where $\vec n_i$ is a unit vector in the direction of the photon $i$ 
and ${\vec n}^\pm$ are the unit vectors in the directions of the 
positively- and negatively-charged leptons.  For each photon, we introduce 
a partition of unity  
\be
1 = 
\frac{\Delta_i^+}{\Delta_i^+ + \Delta_i^-} 
+ 
\frac{\Delta_i^-}{\Delta_i^+ + \Delta_i^-}
= \rho_i^+ + \rho_i^-,
\ee
and obtain
\be
1 = \prod \limits_{i=1}^{2}(\rho_i^+ + \rho_i^-)
= \rho_1^+ \rho_2^+ + \rho_1^- \rho_2^- 
+ \rho_1^+ \rho_2^- + \rho_1^- \rho_2^+.
\ee
We introduce $\rho_i^a \rho_j^b = \delta^{-a, -b}_{ij}$, and re-write 
the previous equation as 
\be
1 = \delta_{12}^{--} + \delta_{12}^{++} 
+ \delta^{-+}_{12} + \delta^{+-}_{12}.
\label{eq1}
\ee 
Each of the contributions on the right-hand side of Eq.~(\ref{eq1}) defines 
a primary sector whose phase space we must parameterize separately. 
The superscripts of each $\delta$ indicate  the potentially singular 
collinear  directions for each of the two photons.  For example, in the primary sector labeled by 
$\delta_{12}^{--}$ both photons can become collinear to the electron.  In the sector 
labeled by  $\delta^{+-}_{12}$ only one photon can be collinear to the electron, while the other can become collinear to the positron.  
Using Eq.~(\ref{eq1}), we decompose the phase-space as 
\be
{\rm dLips}_{e+e-\gamma_1 \gamma_2} 
= \sum _{a,b=\pm} {\rm dLips}_{e+e-\gamma_1 \gamma_2}^{ab}, 
\ee
where 
\be
\begin{split}
& {\rm dLips}_{e+e-\gamma_1 \gamma_2}^{ab} 
= \frac{1}{2!} \int [{\rm d} p_-] [{\rm d}p_+][{\rm d}p_1]
[{\rm d} p_2] 
\\
& \times  (2\pi)^d \delta^d(p_Z -p_--p_+-p_1 - p_2)
\delta_{12}^{ab},
\end{split} 
\label{eq2a}
\ee
For convenience, we introduce the short-hand notation 
$[dp] = {\rm d}^{d-1} \vec p/(2p_0 (2\pi)^{d-1})$, where $d = 4-2\epsilon$ is the 
dimensionality of space-time.  The overall $1/2!$ is the symmetry factor for the two final-state photons.

Out of the four primary sectors, two sectors contain
triple-collinear singularities where both photon momenta are collinear to either
the electron or positron momentum.  The other two sectors contain double-collinear singularities, where the momentum of one photon is collinear to the 
electron momentum while the other photon is collinear to the 
positron.  It is sufficient 
to understand one triple-collinear and one double-collinear partition.  The 
remaining primary sectors are obtained by a simple re-labeling of the final-state particles.  When discussing phase-space parameterizations in the 
relevant sectors, we make use of the fact that the electron and positron 
cannot develop soft singularities, and that in the process 
$Z \to e^+e^- \gamma_1 \gamma_2$,  the kinematic configuration 
where the electron 
and positron momenta are collinear  is non-singular. If such singularities could occur, only an additional partitioning would be required to handle them.

\subsection{The triple collinear sector}
\label{sec:triplec}

\subsubsection{Phase-space parameterization}
We consider the $\delta_{12}^{--}$ primary sector, where the photons and the electron can 
develop collinear singularities. We must also consider soft singularities that appear 
when the energies of one or both photons vanish. 
Our discussion of the phase-space parameterization 
closely follows Refs.~\cite{Czakon:2010td,Czakon:2011ve}.

We first explain how the energies of 
the two photons are parameterized.
We denote the sum of the four-momenta 
of the electron and positron by $Q = p_+ +  p_-$.  Momentum conservation 
implies $p_Z - p_1 - p_2 = Q$ and $0 < Q^2< m_Z^2$. We write 
$m_Z^2 - Q^2 = \Delta m_Z^2$, $0 < \Delta  < 1$.  
Squaring the momentum conservation equation, we find 
\be
m_Z^2 - 2m_Z(E_1 + E_2) + 2E_1 E_1 (1 - \vec n_1 \cdot \vec n_2) = Q^2,
\label{eq3}
\ee
where $E_{1,2}$ are the energies of the two photons and 
$\vec n_{1,2}$ are three-dimensional unit vectors along their momenta.  We parameterize 
the photon energies by $E_i = \xi_i m_Z/2$, and the relative angle between 
them by $\eta_{12} = (1- \vec n_1 \cdot 
\vec n_2)/2$. Solving for $\xi_1$ 
or $\xi_2$ in Eq.~(\ref{eq3}) then yields 
\be
\xi_1 = \frac{\Delta - \xi_2}{1-\xi_2 \eta_{12}}
~~~~{\rm or}~~~~\xi_2 = \frac{\Delta - \xi_1}{1-\xi_1 \eta_{12}}.
\ee

We can remove the symmetry factor in Eq.~(\ref{eq2a}) by requiring that 
$E_1 > E_2$ and by using the fact that the matrix element is symmetric 
under the interchange of $\gamma_1$ and $\gamma_2$.  We obtain 
\be
\begin{split}
& {\rm dLips}_{e+e-\gamma_1 \gamma_2}^{--} 
=(2\pi)^d  \int [{\rm d} p_-] [{\rm d}p_+][{\rm d}p_1]
[{\rm d} p_2] \delta_{12}^{--}  \times 
\\
& \delta^d(p_Z -p_--p_+-p_1 - p_2)
\theta( \xi_1 \xi_{\rm max}(\xi_1) - \xi_2),
\label{eq2}
\end{split}
\ee
where 
\be
\xi_{\rm max}(\xi_1) = {\rm min} 
\left [ 1, \frac{1 - \xi_1}{\xi_1 ( 1 - \xi_1 \eta_{12})} \right ].
\ee
We decompose the four-particle phase-space into  ``regular'' and 
``singular''  phase-spaces:
\be
{\rm dLips}_{e+e-\gamma_1 \gamma_2}^{--} =
{\rm dLips}_{\rm reg} \; \times 
{\rm dLips}_{\rm sing}^{--}, 
\ee
where 
\be
{\rm dLips}_{\rm sing}^{--} = 
[{\rm d}p_1][{\rm d}p_2] 
\;\delta_{12}^{--} \; \theta( \xi_1 \xi_{\rm max}(\xi_1) - \xi_2),
\ee
and 
\be
{\rm d} {\rm Lips}_{\rm reg} 
= [{\rm d} p_-][{\rm d} p_+] (2\pi)^d \delta^{(d)}(Q - p_+ - p_-). 
\ee

We begin with a discussion of  the singular phase-space. We note that since the 
$Z$-boson decays at rest and there are only four particles in the final 
state, we can choose the momenta of any three particles to be in the 
four-dimensional space, without any $(d-4)$-dimensional components. The 
three-momentum  of the fourth particle is determined by momentum 
conservation, and is also in the four-dimensional space.  To have a simple parametrization,
we choose the  direction of the electron momentum to be the 
$z$-axis. Then, 
$p_- =  E_{-} (1,\vec n_-)$, $\vec n_- = (0,0,1)$ 
 and $p_{1,2} = E_{1,2} \left (1, \vec n_1 \right )$,
where $\vec n_1 = (\sin \theta_1, 0, \cos \theta_1 ) $ and 
$\vec n_2 = (\sin \theta_2 \cos \varphi, \sin \theta_2 \sin \varphi, \cos \theta_2 )$.
We also introduce the following notation for the scalar product of $\vec n_-$ with 
$\vec n_{1,2}$: 
\be
\eta_{1,2} =  \frac{1 - \vec n_- \cdot \vec n_{1,2}    }{2}. 
\ee
We then find the scalar products 
\be
\begin{split} 
2 p_i \cdot p_- = 2 E_- m_Z \xi_i \eta_{i},\;\;\;\
2 p_1 \cdot p_2 = m_Z^2 \xi_1 \xi_2 \eta_{12}.
\end{split} 
\ee
We now write the parametrization of the singular phase-space using 
the angular variables  just introduced: 
\be
\begin{split}
& {\rm dLips}_{\rm sing}^{--} = 
\frac{\delta_{12}^{--} \theta(\xi_1 - \xi_2)}{64 \pi^2}\left (\frac{m_Z}{2\pi}   \right )^{2d-4} 
\\
& 
\times {\rm d}\Omega_1^{(d-2)} {\rm d} \Omega_2^{(d-3)} \; 
{\rm d} \xi_1 \; \xi_1^{1-2\ep} \;
{\rm d} \xi_2 \; \xi_2^{1-2\ep}\;
\\
& 
\times {\rm d} \eta_1 \left [ \eta_1 (1-\eta_1) \right ]^{-\ep} 
{\rm d} \eta_2 \left [ \eta_2 (1-\eta_2) \right ]^{-\ep} 
\\
& \times {\rm d} \cos \varphi \left ( 1 - \cos^2 \varphi \right )^{-1/2-\ep}. 
\end{split}
\label{eqpsreg2}
\ee
The goal is to rewrite the integration over $\cos \varphi$ in a way 
that makes the factorization of singularities manifest. To this 
end, we introduce the variable $\kappa$: 
\be
\kappa = \frac{\left ( 1- \cos ( \theta_1 - \theta_2 ) \right )
\left ( 1 + \cos \varphi \right ) }{
2 \left ( 1 - \cos (\theta_1 - \theta_2) 
+ ( 1- \cos \varphi) \sin \theta_1 \sin \theta_2 \right )}.
\ee
Because $\cos \varphi$ can be used to parameterize $\eta_{12}$, we will need a 
relationship between $\eta_{12}$ and $\kappa$.   It can be easily derived 
by solving the above equation for $\cos \varphi$ and then using the solution 
in the expression for $\eta_{12}$.   We find
\be
\eta_{12} = \frac{(\eta_1 - \eta_2)^2}{{\tilde N}(\eta_1,\eta_2,\kappa)},
\label{eq4}
\ee
where 
\be
\begin{split} 
{\tilde N}(\eta_1,\eta_2,\kappa) & =  \eta_1 + \eta_2 - 2 \eta_1 \eta_2
\\
& -2 (1-2\kappa) \sqrt{\eta_1 \eta_2 (1-\eta_1)(1-\eta_2)}.
\end{split} 
\ee

Finally, we need the Jacobian for the $\varphi \to \kappa$ variable 
transformation, and a simple expression for $\sin^2\varphi = 
1- \cos^2 \varphi$.  The relevant 
equations are 
\be
\frac{{\rm d} \cos \varphi}{{\rm d} \kappa } = 
\frac{2 \eta_{12}^2}{(\eta_1 - \eta_2)^2},\;\;\;\;
1 - \cos^2 \varphi = \frac{4 \kappa (1-\kappa) \eta_{12}^2}{(\eta_1 - \eta_2)^2}.
\ee
We can now change  variables $\varphi \to \kappa$ 
in Eq.~(\ref{eqpsreg2})  for the singular 
phase-space. 
We find 
\be
\begin{split}
& {\rm dLips}_{\rm sing}^{--} = 
\frac{ \delta_{12}^{--} \theta(\xi_1 - \xi_2)
m_Z^{2d-4}}{2^{4+2\ep} (2\pi)^{2d-2}}
\\
& \times {\rm d}\Omega_1^{(d-2)} {\rm d} \Omega_2^{(d-3)} \; 
{\rm d} \xi_1 \; \xi_1^{1-2\ep} \;
{\rm d} \xi_2 \; \xi_2^{1-2\ep}\;
\\
& \times {\rm d} \eta_1 \left [ \eta_1 (1-\eta_1) \right ]^{-\ep} 
{\rm d} \eta_2 \left [ \eta_2 (1-\eta_2) \right ]^{-\ep}
\\
& \times {\rm d} \kappa \left ( \kappa (1 - \kappa) \right )^{-1/2-\ep} 
\frac{\eta_{12}^{1-2\ep}}{|\eta_1 - \eta_2|^{1-2\ep}}. 
\end{split}
\label{eq_21_new}
\ee
As we will see later, Eq.~(\ref{eq_21_new}) gives us the singular 
phase space in a form that is  convenient for the extraction 
of singularities.  

Next, we discuss the regular phase space. We write it as 
\be
\begin{split}
{\rm d} {\rm Lips}_{\rm reg} 
& = [{\rm d} p_-][{\rm d} p_+] (2\pi)^d \delta^{(d)}(Q - p_+ - p_-) 
\\
& = \frac{{\rm d} \Omega_{e-}^{(d-1)}}{2 (2\pi)^{d-2}}
\frac{E_-^{1-2\ep}}{2(Q_0 - \vec Q \cdot \vec n_- )},
\label{eqpsreg}
\end{split}
\ee
where $E_- = Q^2/[2(Q_0 - \vec Q \cdot \vec n_-)]$ is the electron energy.  It is important to understand which elements 
of the calculation can be simplified by setting the number of space-time 
dimensions to four, $d \to 4$.  In the context of the phase-space 
discussion, we factor out the leading order phase-space for
$Z \to e^+e^-$ in Eq.~(\ref{eqpsreg}) and treat is as four-dimensional. 
Everything else in Eq.~(\ref{eqpsreg}) is treated with 
exact $\ep$-dependence.
The leading-order phase-space that we use in what follows reads
\be
{\rm d} {\rm Lips}_{Z \to e^+e^-} 
= \frac{{\rm d} \Omega^{(d-1)}}{8(2\pi)^{d-2}}
\left ( \frac{m_Z}{2} \right )^{-2\ep}
\to \frac{{\rm d} \cos \theta {\rm d} \varphi }{32 \pi^2}.
\label{eq4d}
\ee
The regular phase-space becomes 
\be
{\rm d} {\rm Lips}_{\rm reg} =
{\rm d} {\rm Lips}_{Z \to e^+e^-} 
\frac{2 E_-}{(Q_0 - \vec Q \cdot \vec n_- )}
\left ( \frac{2 E_-}{m_Z} \right )^{-2\ep},
\ee
where 
${\rm d}{\rm Lips}_{Z \to e^+e^-}$ is taken in four dimensions, 
as in Eq.~(\ref{eq4d}).

With the explicit parametrization of the phase space at hand, 
we are ready to discuss 
how the momenta of the final-state particles are generated.  We follow the simple procedure described below:
\begin{enumerate} 
\item first, 
we use $\xi_{1,2},\eta_{1,2},\kappa$ to generate the energies and momenta 
of the two photons;
\item second, we calculate $Q = p_Z - p_1 - p_2$;
\item third, we obtain $E_- = Q^2/(2(Q_0 - \vec Q \cdot \vec n_-))$, assuming 
$\vec n_-$ is along the $z$-axis;  
\item finally, the four-momentum 
of an electron is taken as $p_- = E_- (1,0,0,1)$, 
and the four-momentum of the positron is calculated by $p_+ = p_Z - p_1 - p_2 - p_-$.
\end{enumerate}

The remaining issue is the extraction of singularities from 
the matrix element of $Z \to e^+e^-\gamma_1 \gamma_2$ in the $\delta_{12}^{--}$ 
sector. The matrix element is 
singular if either photon is soft, and also if either photon momentum is collinear to the electron 
momentum.  By analyzing the potentially 
singular denominators, it is straightforward to find that there are three 
sectors to consider. These sectors 
are identified by  changes of variables 
$( \xi,\eta,\kappa) \to \{x_{i=1..5} \}$ that we make in order 
to factor out all singularities from the matrix element. The three 
sectors are\footnote{We  use the notation 
$x_{\rm max}$ for the {\it function} $\xi_{\rm max}(x_1)$}: 
\begin{enumerate}
\item ${\it S}_1^{--}$, where $\xi_1 = x_1$, $\xi_2 = x_{\rm max} x_2 x_1$, 
$ \eta_1 = x_3$, $\eta_2 = x_4 x_3$, $\kappa = x_5$;
\item ${\it S}_2^{--}$, where $\xi_1 = x_1$, $\xi_2 = x_{\rm max} x_2 x_4 x_1$, 
$ \eta_1 = x_3 x_4 $, $\eta_2 = x_3$, $\kappa = x_5$;
\item ${\it S}_3^{--}$, where 
$ \xi_1 = x_1$, $\xi_2 = x_{\rm max} x_2 x_1$, 
$\eta_1 = x_2 x_3 x_4$, $\eta_2 = x_3$, $\kappa = x_5$.
\end{enumerate} 
For each of the sectors ${\it S}_i$ we must express the phase-space 
through the new variables and find the singular limits of the amplitudes. 
We illustrate how this is accomplished for the sector $S_1^{--}$.  The remaining two sectors are handled in a 
similar fashion.

\subsubsection{The sector $S_1^{--}$}

For the sector $S_1^{--}$, we write the phase-space in the following form 
\be
\begin{split} 
& {\rm dLips}^{--}_{S_1} = {\underline {\rm dLips}}^{--}_{~S_1} 
\left [ x_1^{4} x_2^2 x_3^2 x_4 m_Z^2 \delta_{12}^{--}\right ],
\\
& {\underline {\rm dLips}}^{--}_{~S_1} = {\rm d Norm} \;
{\rm PS}_w  \; \left ( {\rm PS} \right )^{-\ep}
\\
& \times \frac{{\rm d} x_1}{x_1^{1+4\ep}}
\frac{{\rm d} x_2}{x_2^{1+2\ep}}
\frac{{\rm d} x_3}{x_3^{1+2\ep}}
\frac{{\rm d} x_4}{x_4^{1+\ep}}
\frac{{\rm d} \kappa} {\pi (\kappa (1- \kappa))^{1/2}},
\end{split}
\label{eq21a}
\ee
where 
\be
\label{eq_dnorm}
\begin{split} 
& {\rm d Norm}  =  {\rm d} {\rm Lips}_{Z \to e^+e^-}
\frac{\Gamma(1+\ep)^2 m_Z^{2d-6}}{(4\pi)^d} {\cal B}_\ep^{\rm RR},
\\
& {\cal B}_\ep^{\rm RR}  =  1 
- \frac{\pi^2}{2}\ep^2 - 2 \zeta_3 \ep^3 + \frac{3 \pi^4}{40} \ep^4 
+{\cal O}(\ep^5),
\end{split}
\ee
and the normalization factors are given by 
\be
\begin{split}
{\rm PS}_w  & =   \frac{x_{\rm max}^2(1-x_4)}{N(x_3,x_4,x_5)}
\frac{2 E_-}{(Q_0 - \vec Q \cdot \vec n_- )},\\
{\rm PS} & = 16 \left [ \frac{(1-x_4)}{N(x_3,x_4,x_5)} \right ]^2
(1-x_3)(1-x_4 x_3) 
\\
& \times [\kappa(1-\kappa)] x_{\rm max}^{2}
\left ( \frac{2E_-}{m_Z} \right )^2.
\label{eqpsep}
\end{split}
\ee
Note that the normalization of the various pieces is chosen so that the $m_Z$-dependent factor in ${\rm dNorm}$ can be factored 
out entirely.  This will be true for both the two-loop virtual and 
real-virtual contributions. The factor of $m_Z^2$ 
in the square brackets in Eq.~(\ref{eq21a}) is present to make the product 
of the square bracket and the amplitude 
squared for $Z \to e^+e^- \gamma \gamma$ dimensionless.
Note also that ${\rm dNorm}$ in Eq.~(\ref{eq_dnorm}) is the same 
for all sectors, and that we  will be using it in equations for other 
sectors below.  The function $N(x_3,x_4,x_5)$ appears in the equation  for $\eta_{12}$ 
after it is expressed through the new variables. It reads 
\be
\begin{split}
 N(x_3,x_4,x_5) &  =  1+x_4 - 2x_3 x_4 
\\
& - 2(1-2x_5)\sqrt{x_4 (1-x_3)(1-x_3 x_4)}.
\label{eqnx3}
\end{split} 
\ee
The momenta of all the particles are written through $x$-variables 
in the following way:
\be
\label{eq654}
\begin{split}
& E_1 = \frac{m_Z}{2}x_1,\;\;\;
        E_2 = \frac{m_Z}{2}x_1 x_2 x_{\rm max},
\\
& \eta_{12} = \frac{x_3 (1-x_4)^2}{N(x_3,x_4,x_5)},
\\
&       \cos \theta_1 = 1 - 2 x_3, \;\; \
\cos \theta_2 = 1  - 2 x_3 x_4.
\end{split} 
\ee
Angles $\theta_{1,2}$ are the polar angles for the two photons.  We take $\sin \theta_{1,2}$ to be positive-definite.
We choose photon $\gamma_1$ to be in the $x-z$ 
plane, so that $\varphi_1 = 0$. The azimuthal angle of the photon 
$\varphi_2$ is   calculated to be
\be
\sin \varphi_2 = \frac{\sqrt{4 x_5 (1-x_5)} (1-x_4)}{N(x_3,x_4,x_5)},\;
\cos \varphi_2 = \lambda \sqrt{1 - \sin \varphi_2^2},
\label{eq_sinphi}
\ee
where  
$
\lambda  = {\rm sgn}( (1- \cos \theta_1 \cos \theta_2) - 2 \eta_{12}).
$
Finally, the momenta of the two photons are given by 
\be
\label{eq655}
\begin{split}
& p_{1} = E_1 (1, \sin \theta_1,0,\cos \theta_1 ),\;\;\;\;
\\
& p_{2} = 
E_2 ( 1,\sin \theta_2 \cos \varphi ,\sin \theta_2 \sin \varphi,\cos \theta_2 ).
\end{split} 
\ee
The momentum of the electron is $p_{-} = E_-(1,0,0,1)$, and 
the positron four-momentum is obtained from momentum conservation.

With the parameterization of the phase-space at hand, 
we can discuss extraction of 
singularities from  the matrix element.
To this end, 
the factor in square brackets in Eq.~(\ref{eq21a}) is  combined 
with the amplitude squared.  This 
should give a finite expression in all singular 
limits. We introduce the {\it regular} function 
\be
F_1(\{x_{i=1..5} \}) = 
\left [ x_1^{4} x_2^2 x_3^2 x_4 m_Z^2 \delta_{12}^{--}\right ] 
|{\cal M}_{Z \to e^+e^-\gamma\gamma}|^2,
\ee
A calculation of $Z \to e^+e^- \gamma \gamma$ contribution 
to $Z$-decay 
rate involves integration of the function $F_1$ over the phase-space 
in Eq.~(\ref{eq21a}):
\be
\int {\rm d} {\underline {\rm Lips}}_{~S_1}^{--} F_1(x_1,x_2,x_3,x_4,x_5).
\label{eqint1}
\ee
Because of the structure of ${\rm d}{\underline {\rm Lips}}_{~S_1}^{--}$ in Eq.~(\ref{eq21a}),  the integrand is 
singular if one of the integration variables $x_{i}$, $i=1,..4$ vanishes. 
Such singularities can be extracted by writing 
\be
x_i^{-1-n_i\ep} = -\frac{1}{n_i \ep} \delta(x)
+ \left [ \frac{1}{x_i} \right ]_+ -n\ep
\left [ \frac{\ln x_i }{x_i} \right ]_+ +{\cal O}(\ep^2)
\ee
for all of the singular variables.   The plus-distributions are 
defined in a standard way:
\be
\int \limits_{0}^{1} {\rm d} x \left [ \frac{1}{x} \right ]_+ 
f(x) = 
\int \limits_{0}^{1} {\rm d} x \; \frac{f(x) - f(0) }{x}.
\label{eqresplus}
\ee

Resolving the plus-distributions as in Eq.~(\ref{eqresplus}), 
we find that the integration 
in Eq.~(\ref{eqint1}) requires knowing $F_1$ 
when one or more of its arguments vanishes. 
The calculation of these limits requires care.  Although the function $F_1$ is regular everywhere, the matrix element 
${\cal M}_{Z \to e^+e^-\gamma\gamma}$ is singular when certain $x$-variables 
vanish. The physical limits that correspond to a set of particular 
$x$-variables vanishing can be deduced 
from the expressions for the momenta given in Eqs.~(\ref{eq654},~\ref{eq655}).   
We describe  the relevant limits below. 

\begin{itemize} 

\item If $x_1 = 0$, the energies of both photons vanish.  This corresponds to the double-soft limit. In the double-soft limit, the QED 
eikonal currents completely factorize and we obtain 
\be
\begin{split} 
& |{\cal M}_{Z \to e^+e^-\gamma \gamma}|^2 \to  
e^4 J_1 J_2 |{\cal M}_{Z \to e^-e^+}|^2,
\label{eqsoft}
\end{split}
\ee
where the square of the eikonal current for the photon $i$ reads
\be
J_i = \frac{2 p_- \cdot p_+}{(p_- \cdot p_i)(p_+ \cdot p_i)}. 
\ee
The scalar products are computed using the explicit parametrization of the momenta 
given in Eqs.(\ref{eq654},~\ref{eq_sinphi},~\ref{eq655}).
We obtain for the  function $F_1$
\be
F_1|_{x_1= 0}  = 
 \frac{16 e^4}{m_Z^2}|{\cal M}_{Z \to e^-e^+}|^2.
\ee

\item  If $x_1 \ne 0$ but $x_2 = 0$, the photon $\gamma_1$ is hard 
and the photon $\gamma_2$ is soft. The matrix element becomes 
\be
|{\cal M}_{Z \to e^+e^- \gamma_1 \gamma_2}|^2 \to 
 e^2 J_2  |{\cal M}_{Z \to e^+e^-\gamma_1}|^2
\ee 
in this limit.  Calculating the function $F_1$ for $x_1 \ne 0, x_2 = 0$, we obtain 
\be
\begin{split} 
F_1|_{x_2 = 0}  & = 
e^2 \frac{4 (p_+ \cdot p_-) x_1^2 x_3 }{E_-E_+ x_{\rm max}^2}
\frac{ ( 1- \vec n_+ \cdot \vec n_1)}{\Delta_{12}}
\\
& \times 
|{\cal M}_{Z \to e^+e^- \gamma_1} |^2,
\label{eqx20}
\end{split} 
\ee
where $\Delta_{12} = \prod \limits_{i=1}^{2} \left ( 1 
- \vec n_i \cdot \vec n_+ - \vec n_i \cdot \vec n_- \right )$.

\item Equation (\ref{eqx20})  develops singularities
when $x_3 \to 0$, in which case 
photon $\gamma_1$ becomes collinear to the electron.  
We do not show helicity labels in what follows because helicity is conserved along massless fermion lines. 
In the collinear limit we find 
\be
|{\cal M}_{Z \to e^+e^- \gamma_1} |^2
\approx \frac{2e^2}{s_{1e}} P_{e\gamma}(\ep,z) |{\cal M}_{Z \to e^+ {\tilde e}^-}|^2,
\ee
where the momentum of ${\tilde e}^-$ is given by the sum $p_{-} + p_{1}$, 
$P_{e\gamma}(\ep,z)$ is the $e \to e + \gamma$ splitting function given in the Appendix, 
and  $s_{1e} = 2 p_- \cdot p_1= 2 E_-m_Z x_1  x_3$.  Upon evaluating $F_1$ in that limit, we obtain 
\be
\begin{split}
 F_1|_{x_2=0,x_3=0} = & 
\frac{16 e^4 x_1}{m_Z E_- x_{\rm max}^2 \Delta_{12}} P_{e\gamma}(\ep,z)
\\
& \times  |{\cal M}_{Z \to e^+{\tilde e}^-} |^2.
\end{split}
\ee 
The fraction of energy carried away by the electron in the $e \to e+\gamma$ 
splitting  is expressed through the variable 
$ z = 1/(  1 + m_Z x_1/(2 E_-))$.

\item We next consider the $x_4 = 0$ limit, which 
corresponds to the photon momentum $p_2$ being collinear to  the electron 
momentum $p_-$. We calculate 
\be
\begin{split}
 F|_{x_4=0} = &
 \frac{e^2 x_1^3 x_2 x_3 \delta_{12}^{--} m_Z}{E_- x_{\rm max}} P_{e\gamma}(\ep,z)
\\
& \times  |{\cal M}_{Z \to e^+ {\tilde e}^- \gamma_1}|^2,
\end{split} 
\ee
where  the momentum of ${\tilde e}^-$ is $p_- + p_2$ and 
$z = 1/(1+ m_Z x_{\rm max} x_2 x_1/(2 E_-))$.

\item Finally, we consider  $x_3 = 0$. This limit corresponds to the triple 
collinear limit, when the momenta of photons $\gamma_1$ and $\gamma_2$ 
are parallel to the electron momentum $p_-$.  In the triple collinear limit 
the matrix element factorizes as
\be
\begin{split} 
& |{\cal M}_{Z \to e^+e^- \gamma_1 \gamma_2}|^2 
= \left ( \frac{2e^2}{s_{12e}} \right )^2 
\\
& \times P_{\ep,e\gamma_1 \gamma_2}(\ep,z_e,z_1,z_2) |
{\cal M}_{Z \to e^+{\tilde e}^-} |^2,
\end{split} 
\ee
where  the momentum of ${\tilde e}^-$ is $p_{-} + p_1 + p_2$ 
and 
the triple splitting function  $P_{e\gamma_1 \gamma_2}$ 
can be found in the Appendix.
The energy fractions in this case are given by 
\be
\begin{split} 
& z_e = \left [ 1 + \frac{m_Z}{2E_-} x_1 (1 + x_{\rm max} x_2 ) \right ]^{-1},
\\
& z_1 = \frac{m_Zx_1}{2 E_-} z_e,\;\;\;
z_2 = \frac{m_Zx_1 x_2 x_{\rm max}}{2 E_-} z_e.
\end{split} 
\ee

To compute $F_1$ at  $x_3 = 0$, we introduce the 
notation 
\be
\begin{split} 
 s_{12e} &  
\approx  2 E_- m_Z x_1 x_3 d_{12e} + {\cal O}(x_3^2),
\\
 d_{12e} & = 1 + x_2 x_{\rm max} x_4 
\\
& + \frac{m_Z}{2 E_-} \frac{x_1 x_2 x_{\rm max}(1-x_4)^2}{N(0,x_4,x_5)}.
\end{split} 
\ee
Using this notation, 
the function $F_1$ is easy to write down. As an example, 
we present an explicit result 
for $F_1(x_1,x_2,0,x_4,x_5)$
in the CDR regularization scheme:
\be
\begin{split} 
& F_1|_{x_3=0} =
\frac{e^4}{E_-^2} |{\cal M}_{Z \to e^+{\tilde e}^-}|^2
\Bigg \{ 
\frac{x_1^2 x_2}{2x_{\rm max}} \left [ 
P_1(\ep,z_e,z_1,z_2) \right.
\\
& \left. + P_2(\ep,z_e,z_2,z_1)  \right ] 
+ \frac{x_1^2 x_2^2 x_4}{d_{12e}} P_2(\ep,z_e,z_1,z_2) 
\\
& + \frac{x_1^2 x_2}{d_{12e} x_{\rm max}} P_2(\ep,z_e,z_2,z_1)
 -(1-\ep)^2 \frac{x_1^2 x_2^2 x_4}{d_{12e}^2} 
 \\
 & \times \left ( x_{\rm max} x_2 x_4 
+ \frac{1}{x_{\rm max} x_2 x_4}  \right ) +2\ep(1-\ep) \frac{x_1^2x_2^2x_4}{d_{12e}^2}
\Bigg \}. 
\end{split} 
\ee
The functions $P_1$ and $P_2$ are presented in the Appendix.

\end{itemize}

The above formulae describe all the QED singular limits in the sector $S_1^{--}$. They can be implemented in a 
computer code in a straightforward way.

\subsection{ The double collinear sector}

We next consider the $\delta_{12}^{-+}$ primary sector.  The singularities in this sector arise when the
photon $\gamma_1$ is collinear to  the electron, and the
photon $\gamma_2$ is collinear to  the positron. 
Soft  singularities for both photons are possible. 

\subsubsection{ The phase-space parametrization}
To develop  a suitable description of the four-particle phase space in this partition, we make use of the momentum parametrization developed 
by Catani and Seymour \cite{Catani:1996vz} 
for next-to-leading order calculations. 
Specifically, we adopt the momentum mapping for final-final 
dipoles, using the language of Ref.~\cite{Catani:1996vz}.
We will also  use the nomenclature employed  
in Ref.~\cite{Catani:1996vz} to describe different particles 
contributing to the dipoles, to make 
clear the connection to the discussion in that reference. 

We parameterize the phase-space for $Z \to e^+e^- \gamma_1 \gamma_2$ 
in  two steps. 
In the first step, we treat 
photon $\gamma_1$ as ``emitted'', the electron as ``emitter'' 
and the positron as ``spectator''. The $3 \to 2$ momentum
mapping in this  case is given in  Ref.~\cite{Catani:1996vz}.  
It is determined by the momentum conservation equation  
\be
\gamma_1 + p_- + p_+ = {\tilde p}_{1-} + {\tilde p}_+,
\ee
and  the relations between old  and new momenta:
\be
\begin{split}
& p_1 = z_1 {\tilde p}_{1-}  + y_1 (1-z_1) {\tilde p}_+ + p_{1,\perp},\\
& p_- = (1-z_1) {\tilde p}_{1-}  + y_1 z_1 {\tilde p}_+ - p_{1,\perp},\\
& p_+ = (1- y_1) {\tilde p}_+.
\end{split}
\label{eqdc1}
\ee
The momenta ${\tilde p}_{1-}$ and ${\tilde p}_+$ are light-like:
${\tilde p}_{1-}^2 = {\tilde p}_+^2 = 0$.  The momentum 
$p_{1,\perp}$ is orthogonal to both of them.   For the momentum  parametrization in 
Eq.~(\ref{eqdc1}), the phase-space reads 
\be
\begin{split} 
& {\rm d} {\rm Lips}(p_-,p_+,p_1,p_2) 
=  \frac{{\rm d} {\rm Lips}( {\tilde p}_{1-},{\tilde p}_+,p_2)}{2!}
\\
& \;\;\;\; \times  \frac{{\rm d} y_1 {\rm d} z_1 
{\rm d}  \Omega_{d-2}^{(1)}}{4 (2\pi)^{d-1}}
\left ( 2 {\tilde p}_{1-} \cdot {\tilde p}_+ \right )^{1-\ep}
\\
& \;\;\;\; \times \left ( 1- y_1 \right )^{1-2\ep} y_1^{-\ep} 
\left ( z_1 (1-z_1) \right )^{-\ep}. 
\end{split}
\label{ps_fac}
\ee

In the second step, 
we apply a similar mapping for the momenta of the ``reduced'' 
reaction $Z \to {\tilde p}_{1-} + {\tilde p}_+ + p_2 $, by considering 
$\gamma_2$ as ``emitted'', ${\tilde e}^+$ as ``emitter'' and ${\tilde e}_{1-}$ 
as `` spectator''.  The momentum conservation equation becomes 
\be
{\tilde p}_{1-} + {\tilde p}_+ + p_2
= {\tilde p}_{+2} + {\tilde {\tilde p}}_{1-},
\ee
and the new momentum parameterizations  read 
\be
\begin{split}
& p_2 = z_2 {\tilde p}_{+2} + y_2 (1-z_2) {\tilde {\tilde p}}_{1-}
+ p_{2,\perp}, \\
& {\tilde p}_+ = (1-z_2) {\tilde p}_{+2} + y_2 z_2  {\tilde {\tilde p}}_{1-}
- p_{2,\perp},\\
& {\tilde p}_{1-} = (1- y_2) {\tilde {\tilde p}}_{1-}.
\label{eq_step2}
\end{split} 
\ee
Continuing with the parametrization of the phase-space 
shown in  Eq.~(\ref{ps_fac}), we obtain 
\be
\begin{split} 
& {\rm d} {\rm Lips}(p_-,p_+,p_1,p_2)  
=  \frac{{\rm d} {\rm Lips}( {\tilde {\tilde p}}_{1-},{\tilde p}_{2+})}{2!} 
\\
& 
  \times 
(m_Z^2)^{2-2\ep} 
\frac{{\rm d} y_1 {\rm d} z_1 {\rm d} \Omega_{d-2}^{(1)}}{ 4(2\pi)^{d-1}}
\;\frac{{\rm d} y_2 {\rm d} z_2 {\rm d} \Omega_{d-2}^{(2)}}{ 4(2\pi)^{d-1}}
\\
& \times y_1^{-\ep} z_1^{-\ep} y_2^{-\ep} z_2^{-\ep} 
(1-y_1)^{1-2\ep} 
\\
& \times (1-y_2)^{2-3\ep}(1-z_2)^{1-2\ep}(1-z_1)^{-\ep}.
\end{split}
\label{eq_ps_fac_fin}
\ee

We can express the  momenta of all 
particles through $y_{1,2},z_{1,2}$. We choose 
${\tilde {\tilde p}}_{1-}, {\tilde p}_{2+}$ to be along the positive 
and negative $z$-axis respectively. The corresponding momenta read 
\be
\begin{split}
& {\tilde {\tilde p}}_{1-} = \frac{m_Z}{2}\left ( 1, 0, 0, 1 \right ),~~
{\tilde p}_{2+}  = \frac{m_Z}{2} \left ( 1, 0, 0, -1 \right ). 
\end{split}
\ee
We choose $p_{2,\perp}$ along the $x$-axis and 
use Eq.~(\ref{eq_step2}) to obtain ${\tilde p}_+$ and $p_2$. We find 
\be
\begin{split} 
 p_2 = \frac{m_Z}{2} & \Big ( 
z_2 + y_2(1-z_2), 2 \sqrt{z_2(1-z_2)y_2}, 
\\
& 0, y_2 (1-z_2) - z_2 \Big ),\\
 {\tilde p}_+ = \frac{m_Z}{2} & 
\Big ( 
(1-z_2 + y_2 z_2), - 2 \sqrt{z_2 (1-z_2) y_2}, 
\\
& 0, y_2 z_2 - (1-z_2) \Big ).
\end{split} 
\label{eq543}
\ee
We can write momenta of  the photon $\gamma_1$ and the electron using the above 
equations.  The only semi-intricate step is the derivation of $p_{1,\perp}$.  We find
\be
\begin{split} 
& p_{1,\perp} =  - m_Z \sqrt{z_1 z_2 (1-z_1) y_1 (1-y_2) y_2} 
\vec n_+^{\varphi}
\\
&  
+ m_Z \sqrt{z_1 (1-z_1) y_1 (1-y_2) (1-z_2) } 
\vec n_\perp^{\varphi},
\end{split} 
\label{eq43}
\ee
where 
$$
\vec n_+^{\varphi} =  \left ( \cos \varphi , 0, 0 , \cos \varphi \right ), 
\;\;\vec n_\perp^{\varphi} = \left ( 0, \cos \varphi, \sin \varphi, 0 \right ).
$$
Using Eq.~(\ref{eq43}), we derive the following expression for the energy of the 
photon $\gamma_1$: 
\be
\begin{split} 
& E_1 = \frac{m_Z}{2} \Big [  z_1 ( 1- y_2) 
+ y_1 ( 1-z_1) ( 1- z_2 + y_2 z_2  ) 
 \\
&   -2 \sqrt{y_1 z_1 (1-z_1) z_2 y_2 (1-y_2)} \cos \varphi
\Big ].
\end{split} 
\label{eqw1}
\ee
The energy of the photon $\gamma_2$ can be read off from Eq.~(\ref{eq543}).

We now rewrite Eq.~(\ref{eq_ps_fac_fin}) by factoring out the Born 
phase-space and several other factors, similar to what has been 
done for the triple-collinear sector.  From the momentum parametrization we see
that, in addition to $y_{1,2},z_{1,2}$, we need the azimuthal angle 
$\varphi$ to describe the phase space.  There are no singularities associated 
with this angle in the double-collinear sector, 
so we take it to be  $\varphi = 2\pi x_5$, $ 0 \le x_5 \le 1$. 
In Eq.~(\ref{eq_ps_fac_fin}) 
we identify 
${\rm d} {\rm Lips}( {\tilde {\tilde p}}_{1-},{\tilde p}_{2+})$
with the leading order phase-space 
and obtain
\be
\begin{split} 
& {\rm d} {\rm Lips}(p_-,p_+,p_1,p_2)
= {\rm dNorm} \; {\rm Ps}_w \; {\rm PS}_{\rm gen}^{-\ep}
\\
& \times {\rm d} y_1 {\rm d} z_1 {\rm d} y_2 {\rm d} z_2 {\rm d} x_5
y_1^{-\ep} z_1^{-\ep} y_2^{-\ep} z_2^{-\ep} m_Z^2,   
\end{split}
\label{ps_fac_fina_2}
\ee
where ${\rm dNorm}$ is given in Eq.~(\ref{eq_dnorm}).  The other factors read
\be
\begin{split}
& {\rm PS}_w = \frac{1}{2} (1-y_1)(1-y_2)^2(1-z_2), 
\\
& {\rm PS}_{\rm gen} = 4 (1-y_1)^2 (1-y_2)^3(1-z_2)^2 
\\
& \times (1-z_1)(1 - \cos^2 \varphi).
\end{split} 
\ee

There are two  scalar products that can become singular in the collinear limits:
\be
\begin{split} 
& 2 p_+ \cdot p_2 = (1-y_1) y_2 m_Z^2,
\\
&
2 p_- \cdot p_1 = y_1 (1-y_2)(1-z_2) m_Z^2.
\end{split}
\ee
Analyzing these scalar products 
and the expressions  for the photon energies in Eqs.(\ref{eq543},\ref{eqw1}), we conclude that 
for the photon $i=1,2$, the soft singularity corresponds 
to $y_i = 0, z_i=0$, while the collinear singularity corresponds to $y_i=0$ 
and $z_i \ne 0$.  We also note that the apparent vanishing of scalar 
products and photon energies at $y_i,z_i = 1$ also  implies  
vanishing of the electron or positron energy, and for this reason does not lead to non-integrable singularities. 

The two singular limits for each photon are factorized with the help 
of the two sectors $y_i < z_i$ and $z_i < y_i$. Since there are two photons, 
we get four sectors altogether. We show below the changes of variables needed to completely factorize singularities in each sector:
\be
\begin{split} 
& S^{-+}_1, \text{where } y_1 = x_1,\;\;z_1 = x_1 x_2, \; y_2 = x_3,\;z_2 = x_3 x_4,
\\
& S^{-+}_2, \text{where } y_1 = x_2 x_1,\;\;z_1 = x_1, \; y_2 = x_3,\;z_2 = x_3 x_4,
\\
& S^{-+}_3, \text{where } y_1 = x_1,\;\;z_1 = x_1 x_2 , \;\;\; 
y_2 = x_3 x_4 \;z_2 = x_3,
\\
& S^{-+}_4, \text{where } y_1 = x_2 x_1,\;\;z_1 = x_1, \; y_2 = x_3 x_4,\;z_2 = x_3.\\
\end{split} 
\ee
We note that in sector $S_1^{-+}$, 
 there are  only soft singularities.
In sectors $S_2^{-+}$, $S_3^{-+}$,  there is a soft  singularity 
for one of the photons, and both soft and collinear singularities 
for the other. In sector $S_4^{-+}$  there are soft 
and collinear singularities for both photons.  The extraction 
of all the limits in all sectors is similar to the triple collinear 
limit that we already discussed.  For illustrative purposes, 
we only discuss the most difficult sector $S_4^{-+}$. 

\subsubsection{Sector $S_4^{-+}$} 

To discuss singular limits in sector $S_4^{-+}$ it is  convenient 
to introduce the following short-hand notation for the photon energies:
\be
E_1 = \frac{m_Z}{2} x_1 \Omega_1,\;
E_2 = \frac{m_Z}{2} x_3 \Omega_2, 
\ee
where 
\be
\begin{split}
 \Omega_1 & = (1 -y_2)+x_2 (1-z_1)(1-z_2+y_2 z_2)  
\\
            & - 2 \sqrt{x_2 (1-z_1) z_2 y_2 (1-y_2)}\cos \varphi,
\\
\Omega_2  & = 1+ x_4 \left (1 - z_2 \right ). \\
\end{split}
\ee
In this sector, singularities occur if any of 
the variables $x_1,x_2,x_3,x_4$ vanishes. To enable extraction of 
singularities, we write the phase space as
\be
\begin{split} 
& {\rm d} {\rm Lips}_{e^+e^-  \gamma_1 \gamma_2}
= {\rm dNorm} \; {\rm Ps }_w \; {\rm PS}^{-\ep}
\\
& \times \frac{{\rm d} x_1}{x_1^{1+2\ep}} 
\frac{{\rm d} x_2}{x_2^{1+\ep}} 
\frac{{\rm d} x_3}{x_3^{1+2\ep}}
\frac{{\rm d} x_4}{x_4^{1+\ep}}
{\rm d} x_5
\\
& \times 
\left [ x_1^2 x_2 x_3^2 x_4 m_Z^2 \delta^{-+}_{12}\right],
\end{split}
\label{ps_fac_fina}
\ee
where ${\rm dNorm}$ is the same as in the triple collinear limit. 
The other factors read
\be
\begin{split}
& {\rm PS}_w = \frac{1}{2} (1-y_1)(1-y_2)^2(1-z_2), \\
& {\rm PS} = 4  (1-y_1)^2 (1-y_2)^3 (1-z_1)\\
& \times (1-z_2)^2 (1 - \cos^2 \varphi).
\end{split} 
\ee
The finite function in this case is given by the product 
of the term in brackets in Eq.~(\ref{ps_fac_fina}) and 
the squared matrix element for $Z \to e^+e^- \gamma_1 \gamma_2$:
\be
\begin{split} 
& F_4(\{x_{i=1,..5}\})= x_1^2 x_2  x_3^2 x_4  \delta_{12}^{-+}
|M_{Z \to e^+e^-\gamma_1\gamma_2}|^2.
\end{split} 
\ee

We now describe several of the singular  limits in this sector. 
\begin{itemize}

\item The double-soft limit 
corresponds to $x_1 = 0$ and $x_3 =0$. In this limit, 
the function $F_4$ evaluates to 
\be
F_4|_{x_1=0,x_3=0} = \frac{16 e^4}{m_Z^2 \Omega_1 \Omega_2}
|{\cal M}_{Z \to e^+ e^-}|^2.
\ee

\item If $x_1 = 0$ and $x_4=0$, the photon $\gamma_1$ is soft and the
photon $\gamma_2$ is collinear to the positron. The function 
$F_4$ reads 
\be
\begin{split}
F_4|_{x_1=0,x_4=0}  & = 
\frac{32 x_3  e^4  }{(1-z_2) m_Z^2  \Omega_1 \Delta_{12}}
\\
& \times P_{e\gamma}(\ep,z) |{\cal M}_{Z \to {\tilde e}^+ e^-}|^2,
\end{split}
\ee
where $z = 1/(1 + E_2/E_+)$ and the ${\tilde e}^+$ momentum 
is $p_+ + p_2$. 

\item If $x_1 = 0$, the photon $\gamma_1$ becomes soft and the function 
$F_4$ reads
\be
\begin{split} 
F_4|_{x_1 = 0} & = 
  \frac{8 x_3^2 x_4 e^2 (p_- \cdot p_+) 
 (1- \vec n_- \cdot \vec n_2)}{(1-y_2) (1-z_2) m_Z E_+ \Omega_1 \Delta_{12}}
 \\
& \times |{\cal M}_{Z \to e^+ e^- \gamma_2} |^2. 
\end{split} 
\ee 

\item A new type of
singular limit corresponds to  $x_2 = 0$ and $x_4 = 0$. This double-collinear limit corresponds to 
photon $\gamma_1$  collinear to the electron and 
photon $\gamma_2$ collinear to the positron.
The function $F_4$ evaluates to 
\be
\begin{split}
& F_4|_{x_2 = 0, x_4 = 0} = 
\frac{16 e^4 x_1 x_3 }{m_Z^2 (1-z_2) \Delta_{12}}
\\
& \times    
P_{e\gamma}(\ep,z_{1}) P_{e\gamma}(\ep,z_{2}) 
 |{\cal M}_{Z \to {\tilde e}^- {\tilde e}^+ } |^2,
\end{split} 
\ee
where $z_{1} = 1/(1+E_1/E_-)$ and $z_{2} = 1/(1+E_2/E_+)$. The 
momentum of 
${\tilde e}^-$ is the sum of the $e^-$ and $\gamma_1$ momenta, 
and the momentum of ${\tilde e}^+$ 
is the sum of the $e^+$ and $\gamma_2$ momenta.

\end{itemize}
The calculation of other limits proceeds along similar lines.  The resulting expressions are again straightforward to 
implement in a computer code.

\section{ Virtual corrections to single photon emission} 
\label{sectrv}

In this Section we discuss the one-loop   corrections to the single photon emission 
process $Z \to e^+e^- \gamma$.  We first explain the partitioning 
of the phase-space and its parametrization. We then discuss 
how to compute singular limits of scattering amplitudes.

Consider the process 
$Z \to e^-e^+ \gamma_1$.  Singularities can arise if the photon 
is either soft or collinear to the electron or positron. To account for the 
collinear divergences, we partition the phase space as
\be
1 = \delta_1^{-} + \delta_1^{+},
\ee
where $\delta_1^{\pm} = \rho_1^{\mp}$, and $\rho_1^{\pm}$ are defined
in Section~\ref{sec:doublereal}.  
We write 
\be
{\rm dLips}_{e+e-\gamma_1} 
= \sum _{a=\pm} {\rm dLips}_{e+e-\gamma_1 }^{a}, 
\ee
where 
\be
\begin{split} 
& {\rm dLips}_{e+e-\gamma_1 }^{a} 
= \int [{\rm d} p_-] [{\rm d}p_+][{\rm d}p_1]
\\
& \times (2\pi)^d \delta^d(p_Z -p_--p_+-p_1)
\; \delta_{1}^{a}.
\end{split}
\label{eqrva}
\ee

In what follows, we  consider the primary sector $\delta_1^{-}$, where a  collinear 
singularity can only occur when momenta of  the electron and 
the photon  become parallel. The other sector $\delta_1^{+}$ 
gives a symmetric contribution that can be analyzed identically. 
We use a parametrization that is similar to the two-photon case. 
For ${\rm dLips}_{e+e-\gamma_1 }^{-}$, we parameterize the 
photon energy as $ E_1 = m_Z\xi_1 /2 $ and the relative 
angle between the photon and the electron as $\cos \theta_1 = 1 - 2 \eta_1$. 
The reference frame is fixed by requiring that the electron momentum 
is along the $z$-axis and that the photon momentum is in the $x-z$ plane. 
Explicitly, we write
\be
\begin{split} 
& p_{-} = E_- \left ( 1,0,0,1 \right ),\;\ 
\\
& p_1 = \frac{m_Z \xi_1}{2} \left (1, 
\sin \theta_1,  0, \cos \theta_1 \right ). 
\end{split} 
\ee
The momentum of the positron is determined from momentum 
conservation,  $p_{+} = p_Z - p_{-} - p_1$.
The energy of the electron is found by first computing the momentum 
$Q = p_Z - p_1$ and then calculating 
$E_{-} = Q^2/[2 ( Q_0 - \vec Q \cdot \vec n_e) ]$. Finally, with this 
parametrization of the momenta, and 
borrowing notation that we already used when discussing 
 the double-real emission, we write the  phase space as 
\be
\begin{split} 
& {\rm d}{\rm Lips}_{e^-e^+\gamma}^{-} 
= {\rm}{\rm Lips}_{Z \to e^+e^-}  
\frac{{\rm d} \Omega_{\gamma^{-}}^{(d-2)}}{(2\pi)^{d-1}}
\\
& \times \frac{m_Z^{2-2\ep} E_-}{2 (Q_0 - \vec Q \cdot \vec n_-)}
\left ( \frac{2 E_{-}}{m_Z} \right )^{-2\ep}
\\
& \times 
\frac{{\rm d} \xi_1 }{\xi_1^{1+2\ep} } \frac{{\rm d} \eta_1}{\eta_1^{1+\ep}}
\left ( 1- \eta_1 \right )^{-\ep} 
\left [ \xi_1^2 \eta_1 \delta_1^{-} \right ].
\label{eqrv3}
\end{split} 
\ee

We use Eq.~(\ref{eqrv3}) to construct a
finite, integrable function when it is combined 
with the squared matrix element.   To this end, we note that 
$\xi_1$ and $\eta_1$ are already suitable variables for the extraction 
of singularities, with $\xi_1$ 
controlling the soft limit and $\eta_1$ controlling 
 the collinear 
limit.  The squared matrix element in the real-virtual case is given by 
the interference of  the tree and one-loop $Z \to e^+e^- \gamma$  amplitudes.
We employ Passarino-Veltman reduction to express the one-loop 
scattering amplitude $Z \to e^+e^- \gamma$ in terms of one-loop 
integrals, and use the QCDloops program~\cite{qcdloops} to compute 
master integrals.

At first, it appears that we must simply repeat what we have done for the 
double-real emission corrections, extracting singularities of the matrix elements 
when $\xi_1$ or $\eta_1$ goes to zero.  However, there is 
a subtlety here.  One-loop amplitudes  
are not rational functions of 
$\xi_1$ and $\eta_1$, 
in contrast to their tree-level counterparts. 
To ameliorate this problem, we note that 
the master integrals which produce singularities 
in the $\delta_1^{-}$   sector
can depend  on $ s_{e1} = 2 p_{-} \cdot p_{1}$ 
raised to a non-integer power. Symbolically,
\be
\begin{split}
& 2 {\rm Re} \left ( 
{\cal M}_{Z \to e^+e^-\gamma}^{(1)} 
{\cal M}_{Z \to e^+e^- \gamma}^{(0)*} \right ) 
= B_1 + B_2 (s_{e1})^{-\ep},
\label{eqref105}
\end{split}
\ee
where $B_{1,2}$ are functions that 
can be Taylor expanded around  the $s_{e1} = 0$ limit.
Since $s_{e1} \sim \xi_1 \eta_1$, the second term in the above 
equation provides additional ${\cal O}(\ep)$ contributions to the exponents 
of singular variables.   Obtaining those exponents correctly
is crucial 
for  constructing a valid  expansion of the real-virtual corrections 
in inverse powers of $\epsilon$.  
Because there are two terms in Eq.~(\ref{eqref105}), 
we must introduce 
two functions $F_{1,2}(\xi_1,\eta_1)$ to parameterize 
the matrix element.  We write
\be
\begin{split}
& ~~\left [ \xi_1^2 \eta_1 \delta_1^{-} \right]
 2 {\rm Re} \left ( {\cal M}^{(1)}_{Z \to e^+e^- \gamma} 
{\cal M}^{(0)*}_{Z \to e^+e^- \gamma} \right)  =
\\
& ~~~~~
F_1(\xi_1,\eta_1) + \xi_1^{-\ep} \eta_1^{-\ep} F_2(\xi_1, \eta_1) 
= F(\xi_1,\eta_1).
\end{split}
\label{eqrv4}
\ee
Away from the singular points $\xi_1=0$ and $\eta_1 = 0$, the function 
$F(\xi_1,\eta_1)$ is obtained by computing one-loop corrections 
to the radiative decay $Z \to e^+e^- \gamma$ using standard 
techniques. In the singular limits, we must distinguish between the two contributions.
Inserting Eq.~(\ref{eqrv4}) into the phase space and performing the
plus-distribution expansion for the $\xi_1$ and $\eta_1$ variables, 
we generate many terms with $\xi_1=0$ and/or 
$\eta_1 = 0$.  We now discuss how to obtain $F_{1,2}$ 
at these singular points.

First, we consider the case $\xi_1 = 0$, which corresponds to the photon 
$\gamma_1$ becoming soft. In that limit, both tree and one-loop 
QED amplitudes factorize into the products of the eikonal current 
and the corresponding amplitudes with the photon removed~\cite{Catani:2000pi}:
\be
\begin{split} 
{\cal M}_{Z \to e^+e^- \gamma}^{(0,1))}  & \to  e 
\left ( \frac{p_{-} \cdot \epsilon_1}{p_- \cdot p_1} 
- \frac{p_{+} \cdot \epsilon_1}{p_+ \cdot p_1} 
\right )
{\cal M}_{Z \to e^+e^-}^{(0,1)},
\end{split} 
\ee
where $\epsilon_1$ is the photon polarization vector. 
Using this result, it is easy to find the limit of the 
function $F(\xi_1,\eta_1)$:
\be
\begin{split}
& F|_{\xi_1 = 0} 
= \lim _{\xi_1 =0} \Bigg [ \xi_1^2 \eta_1 \delta_1^- 
\\
& \times 2 {\rm Re} \Big ( {\cal M}^{(1)}_{Z \to e^+e^- \gamma}  
{\cal M}^{(0)*}_{Z \to e^+e^-\gamma} \Big ) \Bigg ]
\\
& = \frac{4 e^2 }{m_Z^2} 2 {\rm Re}
\left ( {\cal M}_{Z \to e^+e^-}^{(1)} 
{\cal M}_{Z \to e^+e^-}^{(0)*} \right). 
\end{split} 
\ee
Because no terms that behave like $\xi^{-\ep}$ appear in this limit,
we conclude that 
\be
F_1(0,\eta_1) = F(0,\eta_1),\;\;\;\;
F_2(0,\eta_1) = 0.
\ee

The next step is the calculation of the collinear $\eta_1 = 0$ limit.
It is much more  involved. The factorization in this case is given 
in terms of splitting amplitudes~\cite{Kosower:1999rx}:  
\be
\begin{split} 
& {\cal M}_{Z \to e^-e^+\gamma_1}^{(0)} \to 
 {\rm Split}_{e_\lambda^* 
\to e_- \gamma}^{(0)}  {\cal M}_{Z \to e^-e^+}^{(0)},\\
& {\cal M}_{Z \to e^-e^+\gamma_1}^{(1)} \to 
  {\rm Split}_{e_\lambda^* 
\to e_- \gamma}^{(0)}  {\cal M}_{Z \to e^-e^+}^{(1)}
\\
& ~~~~~~~~~+ 
{\rm Split}_{e_\lambda^* 
\to e_- \gamma}^{(1)}  {\cal M}_{Z \to e^-e^+}^{(0)}.
\end{split} 
\label{eqrv10}
\ee
Hence,  the one-loop amplitude factorizes into the one-loop splitting  amplitude 
times the tree photon-less amplitude, and the tree splitting  amplitude times the 
one-loop photon-less amplitude. The relevant splitting 
functions were computed in Ref.~\cite{Kosower:1999rx}.  They are given in terms of ``standard matrix 
elements''.  For  an $e^* \to e_a \gamma_b$ splitting, the situation here, the QED splitting amplitudes are
\be
\begin{split}
& ~~~~~~~{\rm Split}^{(0)} = -\frac{\bar u_a \not\!\epsilon_b u_{e^*}}{s_{ab}}, 
\\
& ~~~~~~{\rm Split}^{(1)} =  -2 \Big ( r_3(z){\rm Split}^{(0)}  
- r_4(z) {\rm Split}^{(2)} \Big ),
\end{split} 
\label{eqrv5} 
\ee
where 
\be
{\rm Split}^{(2)} = 
\frac{2 \bar u_a \not\!k_b u_{e^*} ( k_a \cdot \epsilon_b )}{s_{ab}^2}, 
\ee
and the two functions $r_{3,4}(z)$ parameterize loop contributions 
to the splitting functions.   We must square the splitting functions 
and sum over the polarizations of the  final-state particles. 
We find 
\be
\begin{split}
& {\rm Split}^{(0)} \times {\rm Split}^{(0)} 
\to \frac{2}{s_{ab}} P_{e\gamma}(\ep,z),\\
& {\rm Split}^{(0)} \times {\rm Split}^{(2)} 
\to -\frac{2}{s_{ab}} \frac{z(1+z)}{1-z}.
\label{eqs145}
\end{split} 
\ee
We can now use Eqs.~(\ref{eqrv10},~\ref{eqrv5},~\ref{eqs145}), 
to derive the  collinear $\eta_1 \to 0$ 
limit of the matrix element:
\be
\begin{split} 
&~~~ {\rm Re} \left ( {\cal M}_{Z \to e^+e^- \gamma}^{(0)*} 
{\cal M}_{Z \to e^+e^- \gamma}^{(1)} \right )  \to 
\\
&~~~~ \frac{2 P_{e\gamma}(\ep,z)}{s_{1e}}  {\rm Re} \left ( {\cal M}_{Z \to e^+e^-}^{(0)*} 
{\cal M}_{Z \to e^+e^-}^{(1)}
 \right ) 
\\
&~~~~  - \frac{4}{s_{1e}} 
\Bigg (  P_{e\gamma}(\ep,z) r_{3}(z)
 + \frac{z(1+z)}{1-z} r_4(z) \Bigg ).
\\ 
&
\times 
{\rm Re} \left ( {\cal M}_{Z \to e^+e^-}^{(0)*} 
{\cal M}^{(0)}_{Z \to e^+e^-} \right ), 
\end{split} 
\label{eqrv11}
\ee
where $s_{1e} = 2p_-\cdot p_1$. 
The two functions $r_{3,4}(z)$ are proportional to 
$s_{1e}^{-\ep} = \left ( m_Z^2 \xi \eta (1-z) 
 \right )^{-\ep}$ (see 
Ref.~\cite{Kosower:1999rx}). Consequently,
 the first term in the right hand side of 
Eq.~(\ref{eqrv11}) contributes to the function $F_1(\xi_1,0)$, and the second 
term contributes  to function $F_2(\xi_1,0)$.  We find 
\be
\begin{split} 
& F_1|_{\eta_1 = 0} = \frac{\xi_1 P_{e\gamma}(\ep,z) }{E_- m_Z}
{\rm Re} \left ( 2 {\cal M}_{Z \to e^+e^-}^{(0)*} 
{\cal M}_{Z \to e^+e^-}^{(1)} \right ),
\end{split} 
\ee
where further simplifications are possible since $z = 1- \xi_1$. 
The limit of the function $F_2$ is more complicated. It reads 
\be
\begin{split}
&~~~~ F_2(\xi_1,0) = -2{\rm Re} \left [ {\cal M}_{Z \to e^+e^-}^{(0)} 
{\cal M}_{Z \to e^+e^-}^{(0)} \left  (-z  \right )^{-\ep} \right ]
\\
& ~~~~\times \frac{2 \xi_1}{E_- m_Z} 
\left ( P_{e\gamma}(\ep,z) {\tilde r}_3(z) 
+ \frac{z(1+z)}{1-z} {\tilde r}_4(z) \right ), 
\end{split}
\ee
where the two functions ${\tilde r}_{3,4}(z)$ are re-scaled
versions of $r_{3,4}(z)$ in Ref.~\cite{Kosower:1999rx}. 
They are
\be
\begin{split} 
&  {\tilde r}_3  = -\frac{1}{2} \left ( z f_1(z) - 2 f_2 
+ 
\frac{(1-\delta_R \ep) \ep^2}{(1-\ep)(1-2\ep)}f_2 \right ), \\
&  {\tilde r}_4  = \frac{1}{2} \frac{\ep^2 (1 -  \delta_R \ep) }{(1-\ep)(1-2\ep)} f_2,
\end{split} 
\ee
where $\delta_R = 0$ in the FDH scheme~\cite{Bern:2002zk} and $\delta_R = 1$ in conventional dimensional regularization (CDR).  
The two functions $f_{1,2}$ can be found in 
Ref.~\cite{Kosower:1999rx}. 

Finally, we comment on the construction of the expansion of the real-virtual contribution 
in plus distributions. The key point is that after the expansion is 
performed, we are able to get rid of $F_{1,2}(\xi_1,\eta_1)$ in favor 
of $F(\xi_1,\eta_1)$, for all values of $\xi_1,\eta_1$  except the singular 
ones. At the singular points, we have unambiguous expressions 
for $F_{1,2}$, as shown above. 

\section{Regularization schemes}
\label{sec:reg}

We note that most of the formulae presented in the previous sections do not make reference to a 
particular regularization scheme.  They are valid independently of the scheme. Only the splitting functions, 
the tree-level $Z \to e^+e^- \gamma$  amplitude, and 
the one-loop $Z \to e^+e^- \gamma $ and $Z \to e^+e^-$ amplitudes change upon 
switching the scheme.  As an illustration, consider the various contributions to the triple-collinear 
primary sector in Section~\ref{sec:triplec}.   The double-soft contribution to the function $F_1$ in Eq.~(\ref{eqsoft}) is given by 
the product of the square of eikonal currents and the tree-level matrix element 
for the $Z \to  e^+e^-$ process.  The eikonal current is scheme-independent, while 
if we choose to work with physical four-dimensional polarizations 
of the $Z$-boson, the matrix element for $Z \to e^+e^-$ 
becomes scheme-independent as well.

The real-virtual corrections proceed similarly.  As explained  in 
Section~\ref{sectrv} we require the tree and  one-loop matrix element for 
$Z \to e^+e^- \gamma$, the  one-loop matrix element for $Z \to e^+e^-$, 
and the tree- and one-loop splitting functions for $ e \to e + \gamma$. 
All of these objects are scheme-dependent, but this
scheme-dependence is well-understood. In particular, the scheme-dependence 
of the splitting functions is given in Ref.~\cite{Kosower:1999rx}, while  
the scheme-dependence of the one-loop non-singular amplitudes 
can be found in Ref.~\cite{Signer:2008va}.

Hence, it appears that within the framework 
discussed here any regularization scheme is allowed. The only non-trivial, 
scheme-dependent contribution at NNLO that needs to be computed 
explicitly  is the two-loop virtual corrections. We emphasize that in this framework, no ${\cal O}(\ep)$ 
terms of the double real  emission amplitude $Z \to e^+e^- \gamma \gamma$ 
or real-virtual amplitude $Z \to e^+e^-\gamma$ need to be known through 
higher orders in $\ep$, even in CDR.  It seems at first glance that the ${\cal O}(\ep)$ contribution to the $Z \to e^+e^- \gamma$ 
amplitude is required, since it can hit a $1/\ep$ pole when the other photon has become collinear, leading to a finite contribution.  However, it 
has been suggested recently  that this term cancels when the double-real and real-virtual 
corrections are summed~\cite{arXiv:1107.5131}\footnote{The reason for this cancellation is a  well-understood independence 
of NLO QED corrections to $Z \to e^+e^-\gamma$ on the regularization scheme.}.  We 
have checked this statement by comparing the result from summing the double-real and real-virtual contributions in two different ways: with the full ${\cal O}(\ep)$ contribution retained, and with the ${\cal O}(\ep)$ term instead replaced by its collinear limit.  The sum of double-real and real-virtual is identical in these two cases, indicating that this term does indeed not contribution to the final result for the cross section.  We note that simply dropping the ${\cal O}(\ep)$ term would lead to a mismatch between the squared amplitude and the approximation we use in the collinear limits, causing a divergence in the integration.  It is non-trivial to track exactly how the  ${\cal O}(\ep)$ contribution cancels against similar terms in the collinear splitting functions, but since collinear limits are universal, the replacement that we advocate above appears 
to offer  an easy, practical solution.


\section{Numerical checks} 

To prove the utility of this method, we compute the contributions of double-real, real-virtual 
and virtual corrections to the decay rate of the $Z$-boson into leptons. 
For simplicity, we take the coupling of the $Z$-boson to 
leptons to be vector-like, and ignore all diagrams which contain 
photon vacuum-polarization contributions or its unitary cuts.  We compare separately the double-real 
and real-virtual corrections to the four-and three-particle cuts of the vector-vector correlator, which we obtain using the optical theorem.   We have presented these contributions separately because of the significant numerical cancellations between the double-real radiation, the real-virtual corrections, and the two-loop virtual terms that occur when summing them to obtain the total NNLO correction to the decay rate.   The three- and four- particle cuts of three-loop 
master integrals required for such computation can be found 
in Ref.~\cite{GehrmannDe Ridder:2003bm}.
We present this comparison in the CDR scheme. Note, however, that 
the polarization vectors of the $Z$-boson are not continued to $d$-dimensions,
making the tree decay rate $Z \to e^+e^-$   $\epsilon$-independent. We also 
set $m_Z =1$. We write
\be
\begin{split} 
& \Gamma_{Z \to e^+e^-} = \Gamma_{Z \to e^+e^-}^{(0)} 
\Bigg ( 1 + \frac{3}{4} \frac{\alpha}{\pi}
+ \left ( \frac{\alpha}{\pi} \right )^2 \delta^{(2)}
\Bigg ),
\end{split}
\ee
where $\delta^{(2)}$ is further split into three contributions : 
\be
\delta^{(2)} =  
\delta^{(2)}_{RR} + \delta^{(2)}_{RV} + \delta^{(2)}_{VV}.
\ee

The two-loop virtual correction can trivially be obtained from the known result 
for the two-loop fermion form-factor.  For this reason we do not present it 
here.  From the  analytic computation based on the optical theorem we find 
\be
\begin{split}
\label{eq:gehr} 
& \delta^{(2)}_{RR} = \frac{0.5}{\ep^4} + \frac{1.5}{\ep^3} 
- \frac{1.7246}{\ep^2} - \frac{14.074}{\ep}
- 24.228; 
\\
& \delta^{(2)}_{RV} = -\frac{1}{\ep^4} - \frac{3}{\ep^3} 
+ \frac{3.1794}{\ep^2} + \frac{22.88}{\ep}
+ 32.94. 
\end{split} 
\ee
These results can be compared directly to our computations based on 
the soft and collinear limits of the relevant matrix elements. 
Before we present the corresponding results, we note one complication.  There are interference contributions contained in 
$Z \to e^+e^-e^+e^-$  that can not be disregarded. Typically, these 
interference parts of the four-fermion final state correspond to 
certain cuts of non-planar diagrams and, hence, become part of our 
check.  The four-fermion interference contribution 
 only contains collinear singularities, and can be 
analyzed in the same way as the double-real emission contributions  
discussed in Section II.  Because of the existence of this contribution, 
we split the double-real result into $e^+e^-\gamma \gamma$ and $e^+e^-e^+e^-$ 
final states. From our numerical calculation, we obtain 
\be
\begin{split} 
& \delta^{(2),4e}_{RR}  = -\frac{0.1799}{\ep} - 1.79, 
\\
& \delta^{(2),\gamma}_{RR}  = \frac{0.5}{\ep^4} + \frac{1.5}{\ep^3} 
- \frac{1.726(5)}{\ep^2}
 - \frac{13.94(3)}{\ep}
- 22.61(8),  
\\
& \delta^{(2)}_{RV}  = -\frac{1}{\ep^4} - \frac{3}{\ep^3} 
+ \frac{3.179}{\ep^2} 
 + \frac{22.84}{\ep}
+ 32.97(3).
\end{split} 
\ee
The sum of the two double-real emission corrections agrees with Eq.~(\ref{eq:gehr}), as does the real-virtual contribution, indicating the correctness of the numerical results.

\section{Conclusions}

In this paper we have described in detail a subtraction scheme which enables fully differential calculation at NNLO accuracy.  By combining several ideas present in the literature, including an FKS partitioning of the final-state phase space and sector decomposition, the universal singular limits of amplitudes derived over a decade ago can finally be used to obtain actual physical cross sections.  Our ideas are explained using the simple test case of $Z \to e^+e^-$ as an example.  We discussed how to partition the phase space based on the collinear-singularity structure of the matrix element, and presented the explicit phase-space parameterizations from which the soft and collinear singularities can be extracted as poles in $\ep$ using sector decomposition.  The treatment of the real-virtual corrections is described in a way that generalizes to more complicated processes.  Numerical results that check our techniques were presented.  We have chosen to work in the CDR regularization scheme, although the presented framework remains valid in other schemes such as FDH.  It has been pointed out that difficulties exist when extending FDH to NNLO~\cite{Kilgore:2011ta}.  Although they can be fixed~\cite{arXiv:1106.5520}, with our current understanding the use of CDR imposes no additional technical difficulties, as discussed in Section~\ref{sec:reg}.

One point we wish to emphasize about the result presented here is its generalization to more complicated processes.  As mentioned earlier, one of the problems with earlier sector-decomposition based approaches to NNLO calculations was the need to completely reconsider the phase space and extraction of singularities upon changing the process.  In particular, if one knew the NNLO corrections to  $Z \to e^+e^-$, but wanted to study the NNLO corrections to $Z \to e^+e^- \gamma$, now would have to start from scratch.  That is no longer the case for the framework described here.  Differential $Z$ decay serves as a building block for handling all final-state singularities, as we now describe.

We will consider the real-radiation correction  $Z \to e^+(p_{+})+ e^-(p_{-})+\gamma(p_1)+\gamma(p_2)+\gamma(p_3)$, the most difficult contribution, for definiteness.  Introduce the following partition of phase space:
\begin{equation}
1 = \frac{1}{D}\sum_{(i,j) \in (1,2,3)} \left\{ \delta_{ij,+}+\delta_{ij,-}+\delta_{ij,+-}+\delta_{ij,-+}\right\}.
\end{equation}
Here, $\delta_{ij,k}$ allows $p_i$ and $p_j$ to be soft, but not any other particles; it also only allows collinear singularities when $p_i,p_j,p_k$ are collinear.  $\delta_{ij,kl}$ allows only $p_i$ and $p_j$ to be soft, and also allows only the collinear limits $p_i \parallel p_k$ and $p_j \parallel p_l$.  $D$ is the sum of all $\delta$.  It is simple to construct the appropriate $\delta$ functions, as discussed in Ref.~\cite{Czakon:2010td}.
%
%
Consider the partition with $\delta_{12,+}$ for concreteness.  The contribution of this real-radiation correction to the differential cross section is schematically
\begin{eqnarray}
\label{eq:PS5}
\frac{d\sigma}{d{\cal O}_0} =
\int  {\rm dLips}_{e^+e^-\gamma_1\gamma_2\gamma_3}  |{\cal M}|^2 \delta\left( {\cal O}-{\cal O}_0\right) \frac{\delta_{12,+}}{D}
\end{eqnarray}
where ${\cal O}$ is an observable being studied and ${\cal M}$ is the matrix element.  In this partition, there is no soft or collinear singularity associated with $p_3$.  We should therefore be able to use the phase-space parameterization and singularity extraction described in this paper, which handles the double-unresolved limit of photons $p_1$ and $p_2$.  To make this manifest, rewrite the phase space of 
Eq.~(\ref{eq:PS5}) as
\begin{eqnarray}
{\rm dLips}_{e^+e^-\gamma_1\gamma_2\gamma_3}  &=& ds_{+-12} [dp_3] [dp_{+-12}] {\rm dLips}_{e^+e^-\gamma_1\gamma_2} \nonumber \\
& \times & \delta^{(d)} (p_Z-p_3-p_{+-12}),
\end{eqnarray}
where $s_{+-12}$ is the invariant mass of all final-state particles except the hard photon.  The parameterization of momenta in ${\rm dLips}_{e^+e^-\gamma_1\gamma_2}$ is chosen to be the same as in Section~\ref{sec:triplec}.  The form of $p_3$ in this parameterization is irrelevant; no soft or collinear singularities are associated with this momentum.  We can simply reuse the NNLO results for $Z \to e^+e^-$ to obtain the corrections to $Z \to e^+e^-\gamma$.  Adding additional photons to the final state only increases the number of partitions required.  In this sense, $Z \to e^+e^-$ serves as a building block for extracting final-state singularities from any process.  While we have demonstrated this for only one partition, it follows similarly for the others.

We are excited about the possible applications of these ideas to more phenomenologically interesting processes.  We believe there is significant potential for applying these ideas to the calculation of $2 \to 2$ scattering processes at the LHC, and we look forward to their continued development.

\medskip
\noindent
{\bf Acknowledgments} 
K.M. gratefully acknowledges useful conversations  
with Z.~Kunszt and F.~Caola, and would like 
to thank the KITP at UCSB for hospitality during the work on this paper.
This research is supported by the US DOE under contract
DE-AC02-06CH11357 and the grant DE-FG02-91ER40684, by the NSF under grants PHY-0855365 and PHY05-51164, 
and with funds provided by Northwestern University.
    
\medskip
\section*{Appendix: Splitting functions}
\noindent
We collect here the splitting functions that we employed in 
this computation, in the CDR regularization scheme.  
For the $e \to e+ \gamma$ splitting, we have 
\be
P_{e\gamma}(\ep,z) = \frac{2}{1-z} - (1+z) - \ep (1-z).
\ee
For the $e \to e \gamma_1 \gamma_2$ splitting, we find 
 \cite{Catani:1999ss} 
\be
\begin{split}
P_{e\gamma \gamma}(\ep,z,z_1,z_2) 
& = \frac{s_{12e}^2}{2s_{1e}s_{2e}} P_1(\ep,z_e,z_1,z_2)
\\
&+\frac{s_{12e}}{s_{1e}} P_2(\ep,z_e,z_1,z_2) 
\\
& + (1-\ep) \left [ \ep - (1-\ep) \frac{s_{2e}}{s_{1e}} \right ] 
+ ( 1 \leftrightarrow 2),
\end{split}
\ee
where 
the functions $P_{1,2}$ read
\be
\begin{split} 
 P_{1}(\ep,z_e,z_1,z_2) &= z_e \left ( \frac{1+z_e^2}{z_1 z_2}
-\ep \frac{z_1^2+z_2^2}{z_1 z_2} -\ep (1+\ep) \right ),
\\
 P_{2}(\ep,z_e,z_1,z_2) &= \frac{z_e(1-z_1)+(1-z_1)^3}{z_1 z_2}
+ \ep^2(1+z_e) \\
&- \ep(z_1^2 + z_1 z_2 + z_2^2) \frac{(1-z_2)}{z_1 z_2}.
\end{split} 
\ee

\end{document}